\documentclass[a4paper, conference]{IEEEtran}
\IEEEoverridecommandlockouts
\usepackage[nocompress]{cite}
\usepackage{amsmath,amssymb,amsfonts}
\usepackage{algorithmic}
\usepackage{graphicx}
\usepackage{textcomp}
\usepackage{xcolor}
\usepackage{amssymb}  
\usepackage{algorithm}
\begin{document}
\title{\huge RIS Deployment Optimization with Iterative Detection and Decoding in Multiuser Multiple-Antenna Systems\\
}
\author{\IEEEauthorblockN{Roberto C. G. Porto}
\IEEEauthorblockA{\textit{Centre for Telecommunications Studies} \\
\textit{Pontifical Catholic University of Rio de Janeiro}\\
Rio de Janeiro, Brazil \\
camara@aluno.puc-rio.br}
\and
\IEEEauthorblockN{Rodrigo C. de Lamare}
\IEEEauthorblockA{\textit{Centre for Telecommunications Studies} \\
\textit{Pontifical Catholic University of Rio de Janeiro}\\
Rio de Janeiro, Brazil \\
delamare@puc-rio.br} 
}
\maketitle

\begin{abstract}

This work investigates a Reconfigurable Intelligent Surface (RIS)-assisted uplink system employing iterative detection and decoding (IDD) techniques. We analyze the impact of tuning system parameter tuning for several deployment configurations, including the number of users, access point (AP) antennas, and RIS elements on the IDD performance. Analytical results for both active and passive RIS in a single-input single-output (SISO) scenario demonstrate how deployment choices affect system performance. Numerical simulations confirm the robustness of the RIS-assisted IDD system to variations in these parameters, showing performance gains in certain configurations. Moreover, the findings indicate that the insights derived from SISO analysis extend to multiuser MIMO IDD systems.\\
\end{abstract}

\begin{IEEEkeywords}
 Reconfigurable intelligent surface (RIS), Large-scale multiple-antenna systems, IDD schemes, RIS Deployment, Parameter Tuning.
\end{IEEEkeywords}

\section{Introduction}

Reconfigurable Intelligent Surfaces (RIS) \cite{rista} and multiple-antenna systems \cite{mmimo,wence} have emerged as a promising solution to address critical challenges in sixth-generation (6G) wireless communication systems. A fundamental aspect of optimizing RIS-assisted systems lies in the careful selection of system parameters to maximize overall performance.
Recent studies have demonstrated the potential of integrating RIS with channel coding and iterative processing techniques. These works evaluate the performance of RIS-aided multiuser multiple-input single-output (MU-MISO) systems in typical wireless scenarios, based on the 3GPP standard. Numerical results consistently show notable improvements in key performance metrics, particularly in terms of sum-rate and bit error rate (BER).

In the literature, RIS are often deployed in proximity to either the distributed users or their associated access points (APs) within the network, aiming to enhance overall communication performance \cite{rista}. Several studies have explored the impact of RIS deployment on system capacity and performance, providing valuable insights into optimizing these systems \cite{9359653,9548940,9998527}. A critical design consideration in RIS systems is the number of reflecting elements, which directly influences system performance, channel overhead, and channel estimation complexity \cite{9913356,9452133,9110912}. Achieving a balance between these factors is essential for system efficiency. In \cite{zappone,bjornsson,li,zhao,efrem}, the deployment and the number of elements are jointly addressed in the context of multi-RIS full-duplex communication systems.

In this work, we investigate an RIS-assisted uplink multiuser cellular network employing iterative detection and decoding (IDD) \cite{1494998,7105928,8240730,idd_ris1,idd_ris2}. We introduce a simplified analytical framework that examines the effects of deployment on both passive and active RIS configurations. The analysis is corroborated by simulation results, which demonstrate the influence of various system parameters such as deployment configuration, number of users, RIS elements, and AP antennas—on the performance of IDD systems. To the best of our knowledge, no previous work has comprehensively evaluated the performance of RIS-assisted systems with IDD.



\begin{figure*}
    \centerline{\includegraphics[width=0.95\textwidth]{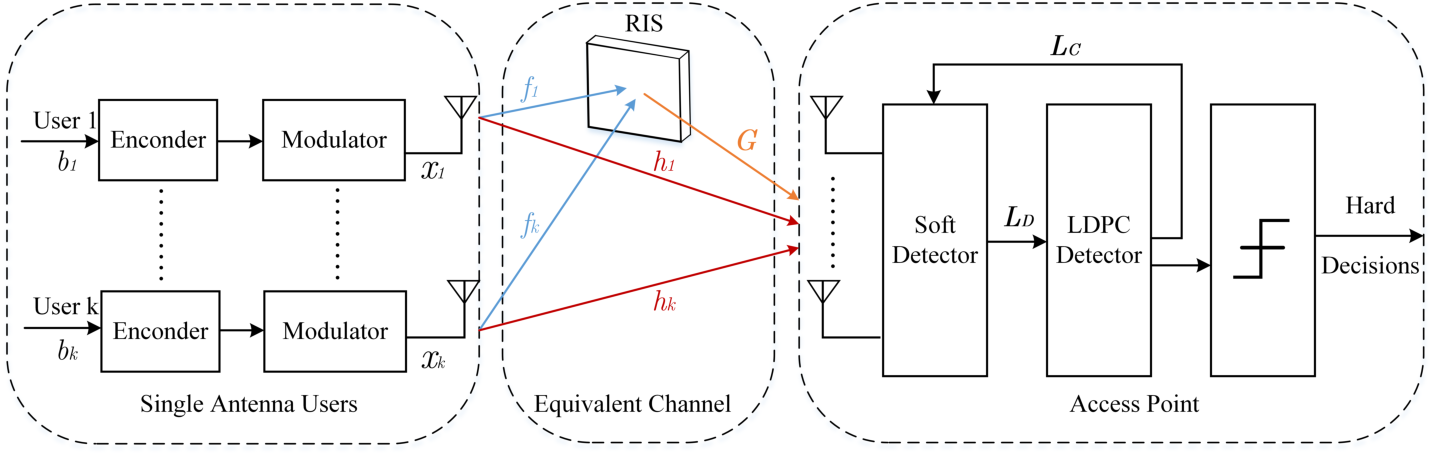}}
    \vspace{-0.725em}
    \caption{System model of an IDD multiuser multiple-antenna system.}
    \label{fig:blockdiagram}
    \vspace{-1em}
\end{figure*}

\section{System Model}

A single-cell MU-MISO system featuring a multiple-antenna access point (AP) assisted by a RIS is considered, as illustrated in Fig. \ref{fig:blockdiagram}. In this configuration, the AP is equipped with $M$ antennas, supporting $K$ users, each equipped with a single antenna. Initially, each user’s information symbols are encoded via individual LDPC channel encoders and subsequently modulated to $x_k$ employing a QPSK modulation scheme. The transmit symbols $x_k$ have zero mean and share the same energy, with $E[|x_k|^2] = \sigma^2_x$. These modulated symbols are then transmitted over block-fading channels.

The RIS is assumed to incorporate $N$ reflecting elements and their reflection coefficients are modeled as a complex vector $\boldsymbol{\varphi} \triangleq [p_1e^{j\theta_1}, \dots, p_ne^{j\theta_N}]^T$, where $\theta_n$ represents the phase shift of the $n$th unit, and $p_n \in \mathbb{R}_+$ denotes the gain factor of the  $n$th unit, for all $n \in \mathcal{N} \triangleq \{ 1, 2, \dots, N\}$. This expression allows for the representation of both passive and active RIS systems. The distinction between the two lies in the fact that in the passive system, $p_n = 1$, for all $n$.

The signal model of an $N$-element RIS is given by a diagonal phase shift matrix $\Phi \in \mathbb{C}^{N\times N}$ defined as $\mathbf{\Phi} \triangleq \text{diag}(\boldsymbol{\varphi})$. We can represent the equivalent channel from the AP to user $k$, which includes both the direct link and the reflected link, denoted as $\mathbf{\bar{h}_k} \in \mathbb{C}^{M \times 1}$, by 
\begin{equation}
    \mathbf{\bar{h}_k} = \mathbf{h_k} + \mathbf{G}\mathbf{\Phi}\mathbf{f_k},
    \label{heff}
\end{equation}
where $\mathbf{h_k} \in \mathbb{C}^{M \times 1}$, $\mathbf{G} \in \mathbb{C}^{M \times N}$, and $\mathbf{f_k} \in \mathbb{C}^{N \times 1}$ represent the communication links from the AP to the user $k$, from the AP to the RIS, and from the RIS to the user $k$, respectively. 

An estimate $\hat{x}_k$ of the transmitted symbol of the $k$th user without soft interference cancellation (SIC) is obtained by applying a linear receive filter $\mathbf{w_k}$ to the received signal $\mathbf{y}$:
\begin{equation}
    \hat{x}_k = \mathbf{w_k}^H\mathbf{y} = \mathbf{w_k}^H\left(\sum_{i=1}^K\mathbf{\bar{h}_k}x_k + \mathbf{G}\mathbf{\Phi}\mathbf{n_v} +\mathbf{n_s}\right),
    \label{detection_estimate_1}
\end{equation}
where the vector $\mathbf{\Phi}\mathbf{n_v}$ denotes the noise introduced and amplified by the active RIS , as described in \cite{9998527}. This noise is characterized as 
$\mathbf{n_v} \sim \mathcal{CN}(\mathbf{0_M}, \sigma_v^2 \mathbf{I_N})$. The amplified noise is subsequently multiplied by the communication links from the RIS and the AP, represented by $\mathbf{G}$. For a passive RIS, $\sigma_v$ is set to zero. The vector $\mathbf{n_s} \sim \mathcal{CN}(\mathbf{0_M}, {\sigma_s^2} \mathbf{I_M})$ represents the static noise. Since the matrix $\mathbf{\Phi}$ is diagonal, the symbol estimate in (\ref{detection_estimate_1}) can also be written in terms of $\boldsymbol{\varphi}$ as given by
\begin{equation}
    \hat{x}_k = \mathbf{w_k}^H(\sum_{i=1}^K\mathbf{h_k}x_k + \sum_{i=1}^K\mathbf{A_k}\boldsymbol{\varphi}x_k + \mathbf{B}\boldsymbol{\varphi} +\mathbf{n_s}),
    \label{detection_estimate_2}
\end{equation}
where $\mathbf{B}=\mathbf{G}\text{diag}(\mathbf{n_v})$ and $\mathbf{A_k} = \mathbf{G} \text{diag}(\mathbf{f_k})$.

\subsection{Enhancing Detection through SIC}

An IDD scheme with a soft detector and LDPC decoding is used to enhance the performance of the system. The soft detector incorporates extrinsic information provided by the LDPC decoder ($\mathbf{L_C}$), and the LDPC decoder incorporates soft information provided by the MIMO detector ($\mathbf{L_D}$), as illustrated in Fig. \ref{fig:blockdiagram}.

Let us denote $\mathbf{\tilde{x}}=[\tilde{x}_1,\dots,\tilde{x}_K]^T$ and
\begin{equation} 
    \tilde {\mathbf {x}}_{k} = \tilde {\mathbf {x}} - \tilde {x}_{k}\mathbf {e}_{k},
\end{equation}
where $e_k$ is a column vector with all zeros, except that the $k$th element is equal to 1. For each user $k$, the interference from the other $K - 1$ users is canceled according to
\begin{equation} 
    \mathbf {y}_{k}=\mathbf {y}-\sum _{j=1,j\neq k}^{K}\tilde {x}_{j}\mathbf {\bar{h}_j}=\mathbf {y}-\mathbf {\bar{H}}\tilde {\mathbf {x}}_{k}. 
\end{equation}
where $\mathbf{\bar{h}_j}$ represents the lines of the matrix $\mathbf {\bar{H}} = [\mathbf {\bar{h}_1}, \dots, \mathbf {\bar{h}_K}]^T$.

A detection estimate $\hat{x}_k^{\rm sic}$ of the transmitted symbol on the $k$th user is obtained by applying a linear filter $\mathbf{w_k}$ to $\mathbf{y_k}$:
\begin{equation}
    \hat{x}_k^{\rm sic} = \mathbf{w_k}^H\mathbf{y_k}
    \label{sic}
\end{equation}
Inspired by prior work on IDD schemes \cite{1494998,8240730,spa,mfsic,mbdf,dfcc,did,1bitidd,listmtc,detmtc,msgamp1,msgamp2,comp,refllr,oclidd}, the soft estimate of the $k$th transmitted symbol is firstly calculated based on the $\mathbf{L_c}$ (extrinsic LLR) provided by the channel decoder from a previous stage to the receiver.

\subsection{Sum-Rate for RIS-MISO systems}

The computation of the sum-rate without the SIC relies on the SINR computed for each user using the MU-MISO transmission model presented in (\ref{detection_estimate_1}). The SINR for the $k$-th user, denoted as $\gamma_k$, is given by

\begin{equation} 
    {\gamma _{k}} = \frac {{{{\left |{ {{{ {\mathbf w}}^{\mathrm{ H}}_{k}}{\mathbf {\bar{h}_k}}} }\right |}^{2}}}\sigma_x ^{2}}{{\sum \nolimits _{j = 1,j \ne k}^{K} {{{\left |{ {{{ {\mathbf w}}^{\mathrm{ H}}_{k}}{\mathbf {\bar{h}_j}}} }\right |}^{2}}\sigma_x ^{2}} + ||{\mathbf w}^{\mathrm{ H}}_{k}\mathbf{G}\mathbf{\Phi}||^2\sigma_v ^{2} +||{\mathbf w}^{\mathrm{ H}}_{k}||^2 \sigma_s ^{2}}}. 
\end{equation}

When considering the scenario with SIC as expressed in (\ref{sic}), the output of the SIC-MMSE filter can be approximated as a complex Gaussian distribution due to the large number of independent variables \cite{idd_ris1,idd_ris2} by:

\begin{equation}
    \hat{x}_k^{\rm sic} = \mu_kx_k + z_k
    \label{eq:approx}
\end{equation}
where $\mu_k$ is the equivalent amplitude of the $k$ user's signal, and the parameter $z_k$ represents the combination of noise and the residual intersymbol interference, which is equivalent to $z_k \sim \mathcal{N}(0,\eta_k^2)$. Using this equation, the SINR at the output of the soft instantaneous MMSE filter can be defined as \cite{8240730}:

\begin{equation}
     {\gamma _{k}^\text{sic}} \triangleq   
     \frac{E[(\hat{x}_{k}^{\text{sic}})^2]}{\text{var}[\hat{x}_{k}^{\text{sic}}]} = 
     \frac{\mu_k^2\sigma_x^2}{\eta_k^2},
\end{equation}
and the corresponding Sum-Rate, denoted as $R_{\mathrm{sum}}$, is then calculated as:
\begin{equation}
    R_{\mathrm{sum}}= \sum \limits _{k = 1}^{K} {\log _{2}\left ({{1 + {\rm SINR_{k}}} }\right)}
    = \sum \limits _{k = 1}^{K} {\log _{2}\left ({{1 + {\gamma _{k}^\text{sic}}} }\right)}.
\end{equation}

\section{Tuning Parameters for Optimal Performance}

In this section, we investigate the selection of key parameters, such as RIS position, the number of users, AP antennas, and RIS elements, and examine their impact on overall system performance in IDD uplink system. 

\subsection{Design of Reflection Parameters for Active and Passive RIS}

\begin{figure}
\vspace{0em}
    \centerline{\includegraphics[width=0.4\textwidth]{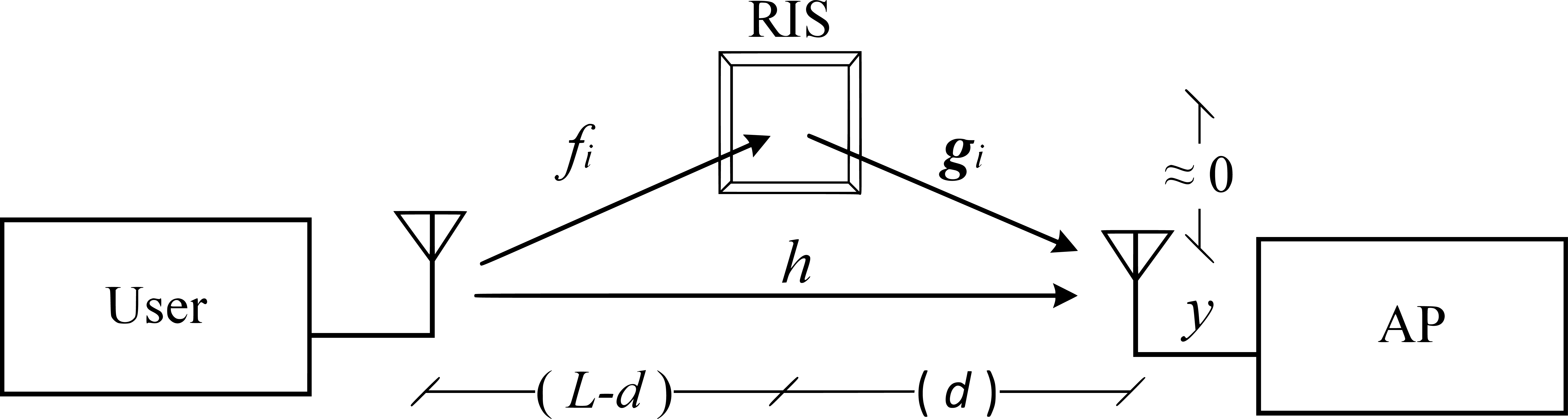}}
    \vspace{0em}
    \caption{Block Diagram of RIS Deployment in SISO Systems}
    \label{fig:deploy}
    \vspace{-1em}
\end{figure}

The optimization of reflection parameters builds upon the work referenced in \cite{idd_ris2}, which focuses on the design of reflection parameters using a minimum mean square error (MMSE) criterion. This approach yields effective results in refining LLRs within an IDD system. Specifically, an MMSE receive filter is employed to facilitate SIC at the receiver. The optimization of reflection parameters for both passive and active RIS is expressed as follows:
\begin{equation} 
\boldsymbol{\varphi}_o = \big[\boldsymbol{\beta}+\frac{\sigma_v^2}{\sigma_x^2}(\mathbf{WG})^H(\mathbf{WG})\big]^{-1}\boldsymbol{\Psi},
\label{eq:phi}
\end{equation}
where {\footnotesize$\boldsymbol{\beta} = \sum^{K}_{i}\mathbf{(WA_i)^H(WA_i)}$, $\boldsymbol{\Psi} = \sum_i^K\mathbf{(WA_i)^H(e_i-W\mathbf{\bar{h}_i})}$} and  $\mathbf{e_i}$ is a column vector with all zeros, except for a one in the ith element. While the norm consistently exhibits convex behavior, the constraint on diagonal elements of the matrix $\mathbf{\Phi}$ does not form a convex set \cite{tds1,tds2}. To address this, we employ a relaxation of this constraint. However, the solution to the problem results in a full-rank matrix $\mathbf{\Phi}$. 

Since we have opted for the relaxation of the constraint for MMSE processing \cite{jidf,idd_ris1,idd_ris2}, this approach involves truncation at the reflection parameters based on whether the RIS is active or passive. For the passive RIS, the constraint is:
\begin{equation}
|[\boldsymbol{\varphi}]_n|=1 , \text{ for } \forall n, \text{ and } \sigma_v^2=0.
\end{equation}
which leads to following truncation: 
\begin{equation}
    \boldsymbol{\varphi_t}^\text{passive} = \frac{[{\varphi}_o]_i}{|[{\varphi}_o]_i|} = e^{j\measuredangle (\boldsymbol{\varphi_0})}.
\end{equation}

For the active RIS, the constraints are:
\begin{equation}
 \sum^K_{i=1} ||\mathbf{\Phi}\mathbf{f}_i||^2\sigma_x^2 +||\boldsymbol{\varphi}||^2\sigma_v^2\leq \text{P}_\text{RIS},  \text{ where } p_n \geq 1,  \text{ for } \forall n.
 \label{eq:active_constrain}
\end{equation} 
and the corresponding truncation is:
\begin{equation} 
    \boldsymbol{\varphi_t}^\text{active} =  \boldsymbol{\varphi_0} .\left( \frac{\text{P}_\text{RIS}}{ \sum^K_{i=1} ||\mathbf{\Phi}\mathbf{f}_i||^2\sigma_x^2 +||\boldsymbol{\varphi}||^2\sigma_v^2}\right)^\frac{1}{2}.
\end{equation}

\subsection{Passive RIS Deployment in SISO Systems}
Due to the multiplicative fading effect, a passive RIS exhibits improved performance when positioned closer to either the access point (AP) or the users, as demonstrated in \cite{9998527}. To illustrate this, we provide a mathematical analysis for a simplified scenario involving a single-antenna AP, one user, and $N$ RIS elements.

Given the system is SISO and does not utilize IDD, the received signal at the AP can be expressed as: 
\begin{equation}
    y = \bar{h} x + n = \left( h + \sum_{i=1}^N g_i f_i \varphi_i \right) x + n. 
\end{equation} 

The RIS deployment strategy is illustrated in Fig. \ref{fig:deploy}. Our goal is to determine the optimal RIS placement by finding the distance $d$ that maximizes the SNR. It is assumed that the vertical distance between the users, AP, and RIS is negligible, leading to the following optimization problem:
\begin{equation} 
    d = \underset{d}{\arg\max} \frac{\mathbf{E} \left[\left| \left( h + \sum_{i=1}^N g_i f_i \varphi_i \right) x\right|^2\right]}{\mathbf{E} \left[|n|^2\right]}, \end{equation} 
where $\mathbf{E}[\cdot]$ denotes the expectation operator.

The coefficients $f_i$, $g_i$, and $h$ correspond to the large-scale fading of the respective channels, with path loss models derived from the 3GPP standard \cite{access2010further}. The path loss is modeled as: 
\begin{equation} 
    \text{A}^z = -(37.3 + 22.0 \log d) \quad \text{(dB)}, 
    \label{3gpp} 
\end{equation} 
where $z$ denotes the corresponding path for channels $f_i$, $g_i$, or $h$. The small-scale fading effects are modeled using Rayleigh fading for all channels, and we assume that the channels are independent.
    
Assuming that $f_i$ and $g_i$ are independent and identically distributed (IID) and independent of $x$, the optimization problem can be rewritten as: 
\begin{equation}
    d = \underset{d}{\arg\max} \frac{(\text{A}^h + \text{A}^g \text{A}^f |\sum_{i=1}^N \varphi_i|^2)\sigma_x^2}{\sigma_n^2}.
\end{equation}

Since the terms dependent on $d$ are $\text{A}^f$ and $\text{A}^g$, the problem can be reformulated as: 
\begin{equation} 
    d = \underset{d}{\arg\max} \text{A}^g \text{A}^f. 
\end{equation}

Substituting the path loss values from (\ref{3gpp}) and assuming the system distances as illustrated in Fig. \ref{fig:deploy}, we obtain: 

\begin{align}
    \begin{split}
        d & = \underset{d}{\arg\max} (10^{-3.73} d^{-2.2})(10^{-3.73} (L-d)^{-2.2})
       \\ & = \underset{d}{\arg\max} (d(L-d))^{-1}
       \\ & = \underset{d}{\arg\min} (d(L-d)) \text{, for } 0 \le  d \le L.
    \end{split}
\end{align}

The minimum value of this function occurs at $d = 0$ (i.e., when the RIS is positioned closer to the AP) or $d = L$ (closer to the user), indicating that the RIS should be placed as close as possible to either the AP or the user for optimal performance.
        
\subsection{Active RIS Deployment in SISO Systems}
The design of active RIS in multi-user scenarios is more complex compared to passive RIS, as it requires a careful balance between user performance and the trade-off between signal amplification and noise. This makes the placement of the active RIS near users or the AP more challenging \cite{kang}. 

Using the same approach as the passive case in the system illustrated in Fig. \ref{fig:deploy}, our objetive is to find the optimal active RIS position that maximizes the SNR at the receiver. In this case we need to take account the dynamic noise from the active RIS, and also the power constrain of the surface expressed in (\ref{eq:active_constrain}). The optimization problem can be written as:

\begin{equation} 
    d = \underset{d}{\arg\max} \frac{\mathbf{E} \left[\left| \left( h + \sum_{i=1}^N g_i f_i \varphi_i \right) x\right|^2\right]}{\mathbf{E} \left[|\sum_{i=1}^N f_i \varphi_iv_i|^2\right]+\mathbf{E} \left[|n|^2\right]}, 
\end{equation}
\begin{align*} 
 {\text{subject to}}\quad |\sum^N_{i=1} f_i \varphi_i|^2\sigma_x^2 +|\sum^N_{i=1} \varphi_i|^2\sigma_v^2\leq \text{P}_\text{RIS}
\end{align*}

Also assuming that the channels have large-scale fading modeled by (\ref{3gpp}), and he small-scale fading effects are modeled using Rayleigh, the optimization problem can be rewritten as: 
\begin{equation} 
    d = \underset{d}{\arg\max} \frac{(\text{A}^h + \text{A}^g \text{A}^f |\sum_{i=1}^N \varphi_i|^2)\sigma_x^2}
    { |\sum_{i=1}^N \varphi_i|^2\text{A}^f\sigma_v^2 + \sigma_n^2}.
    \label{eq:activeObj}
\end{equation} 
\begin{align*} 
 {\text{subject to}}\quad  |\sum^N_{i=1} \varphi_i|^2\text{A}^f\sigma_x^2 +|\sum^N_{i=1} \varphi_i|^2\sigma_v^2\leq \text{P}_\text{RIS}.
\end{align*}

The constraint can be removed by introducing an appropriate substitution. Specifically, it is assumed that the power of the RIS is 10 percent of the user signal power and that all available power is fully utilized. Under this assumption, the constraint can be reformulated as follows:

\begin{equation} 
    |\sum^N_{i=1}  \varphi_i|^2\text{A}^f\sigma_x^2 +|\sum^N_{i=1} \varphi_i|^2\sigma_v^2 =  \frac{\sigma^2_x}{10}.
\end{equation} 

Reorganizing this equation, it is possible to write the summation of the RIS elements as:

\begin{equation} 
\left|\sum^N_{i=1} \varphi_i\right|^2=  \frac{\sigma^2_x}{10(\text{A}^f\sigma_x^2 + \sigma_v^2)}.
\label{eq:constrain}
\end{equation} 

Substituting (\ref{eq:constrain}) to (\ref{eq:active_constrain}):


\begin{align}
    \begin{split}
    d &= \underset{d}{\arg\max} 
    \frac{10\text{A}^h\sigma_x^2\sigma_v^2 + 10\text{A}^h\text{A}^f\sigma_x^4 +  \text{A}^g\text{A}^f\sigma_x^4}
    {\text{A}^f\sigma_v^2\sigma_x^2 +  10\text{A}^f\sigma_x^2\sigma_n^2 + 10 \sigma_n^2\sigma_v^2 }
    \end{split}
    \label{eq:activeObj}
\end{align}

Upon analyzing the numerator, it is observed that the first term remains independent of the distance, the second term increases as the RIS moves closer to the AP, and the third term increases when the RIS is positioned near the AP due to the presence of the factor $\frac{1}{\text{A}^f}$. In contrast, for the denominator, the first two terms decreases when the RIS is closer to the user and increases as the RIS approaches the AP, while the third term is independent of the distance. This allows for the consideration of two extreme cases: when the RIS is positioned near the user and when it is near the AP.

When the RIS is positioned closer to the user, the term $\text{A}^g$ approaches $\text{A}^h$, and $\text{A}^f$ tends towards infinity. Under these conditions, the resulting SNR is given by: 
\begin{equation} 
    \text{SNR}=\frac{11\text{A}^h\sigma_x^2}{\sigma_v^2 + 10\sigma_n^2}. 
\end{equation} 

On the other hand, when the RIS is positioned closer to the AP, the opposite occurs, where $\text{A}^g$ dominates the numerator. This leads to the SNR tending towards infinity, as the contribution from $\text{A}^f$ becomes predominant in the numerator. 

In single-way communication, the optimal scenario is to deploy the active RIS closer to the AP, as this maximizes the received signal strength at the user end. However, in the case of two-way communication, a more balanced strategy would involve placing the RIS at an intermediate position between the AP and the user. This placement helps to balance the performance across both communication directions, ensuring improved efficiency for both uplink and downlink transmissions.

\subsection{Number of users, AP antennas, and RIS elements}

The channel of a RIS-assisted system can be modeled as an equivalent channel, as shown in (\ref{heff}), allowing the system to be viewed as a conventional MIMO system. As a result, performance is expected to improve with the increase in AP antennas and the number of users, up to the limit of the number of AP antennas. Regarding the number of RIS elements, \cite{hzhang} examines the fundamental relationship between the number of reflective elements in RIS and the system sum-rate in multi-user communications. It can be concluded that the sum-rate increases with the number of RIS elements until a threshold, beyond which it becomes upper-bounded. The impact of the IDD system on these parameters will be analyzed in the Results Section. 

\section{
Results}

A short-length regular LDPC code \cite{memd,vfap}  with a block length of $n=512$ and a rate of $R = 1/2$ is considered with QPSK modulation. The channel is assumed to undergo block fading with perfect and estimated channel state information (CSI) at the receiver. The systems operate at a frequency of 5 GHz, with the direct link weakened due to severe obstruction. To characterize the large-scale fading of the channels, path loss models from the 3GPP standard \cite{access2010further} are utilized: ${\mathrm{PL}{w}}= 41.2 + 28.7\log d$ and ${\mathrm{PL}{s}}=37.3 + 22.0\log d$. The path loss model ${\mathrm{PL}{w}}$ is used for the weak AP-user connection, while ${\mathrm{PL}{s}}$ models the strong connections between AP-RIS and users-RIS channels. To incorporate the effects of small-scale fading, the Rayleigh fading channel model is adopted for all channels.

The AP is positioned at (0 m, 0 m), while user locations are randomly distributed within a 5 m radius circle centered at (400 m, 0 m).  The transmit power per user ($P_\text{u}$) is normalized by the code rate, with equal power for all users. For the active RIS scenario, RIS power consumption is limited by the total transmission power ($P_\text{T}$), such that $P_\text{RIS} = 0.1 \times P_\text{T}$ and the power of each user is $P_\text{U} = 0.9 \times P_\text{T} / K$.  Results are presented as total transmission power divided by the number of users ($P_\text{T}/K$). To demonstrate the effectiveness of parameter tuning, the following simulation schemes are considered:
\begin{itemize}
    \item \textbf{Linear MMSE P-RIS:} refers to a system with a linear MMSE receiver and a passive RIS.
    \item \textbf{Linear MMSE A-RIS:} refers to a system employing a linear MMSE receiver with an active RIS.
    \item  \textbf{IDD MMSE P-RIS $\boldsymbol{\tau}$:} refers to a system using a soft MMSE receiver with iterative detection and decoding (IDD) and a passive RIS, optimized over $\tau$ iterations.
    \item \textbf{IDD MMSE A-RIS $\boldsymbol{\tau}$:} refers to a system utilizing a soft MMSE receiver with IDD, combined with an active RIS and optimized over $\tau$ iterations.
\end{itemize}

{RIS Deployment}

To demonstrate the effectiveness of the proposed approach, simulations were conducted for both passive and active RIS configurations, using a fixed value of $P_T/K$. In both cases, the system is configured with $K=12$ users, $M=32$ AP antennas, and $N=64$ RIS elements. The reconfigurable intelligent surface (RIS) is positioned along a horizontal line from (0 m, 10 m) to (400 m, 10 m), where $d$ represents the horizontal distance between the AP and the RIS, denoted as ($d$, 10 m).

\begin{figure}
    \centering
    \includegraphics[width=0.95\linewidth]{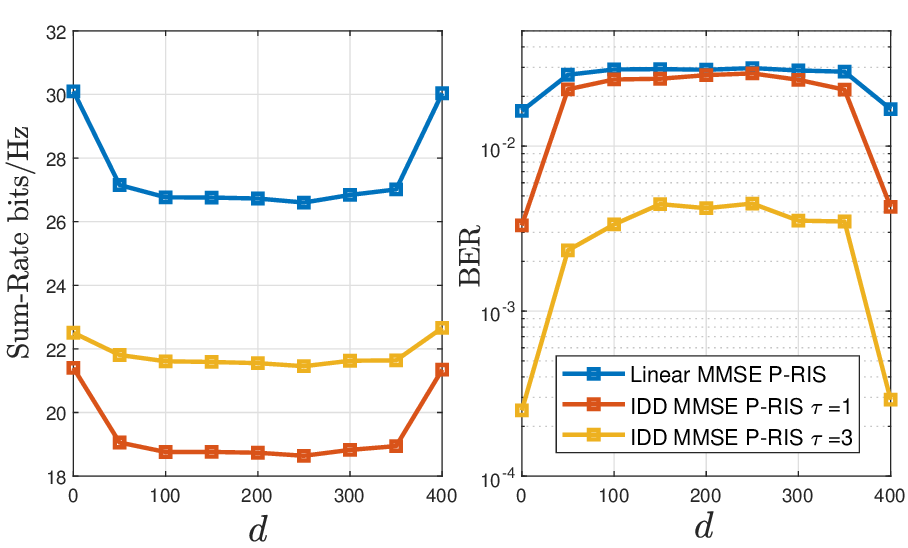}
    \caption{Performance for $K = 12$, $M = 32$, $N = 64$, $\sigma_s^2 = -100$ dBm, $\sigma_v^2 = 0$ dBm and $P_T/K = 6$ dBm.}
    \label{fig:deploy_passive}
    \vspace{-0.5cm}
\end{figure}

\begin{figure}
    \centering
    \includegraphics[width=0.95\linewidth]{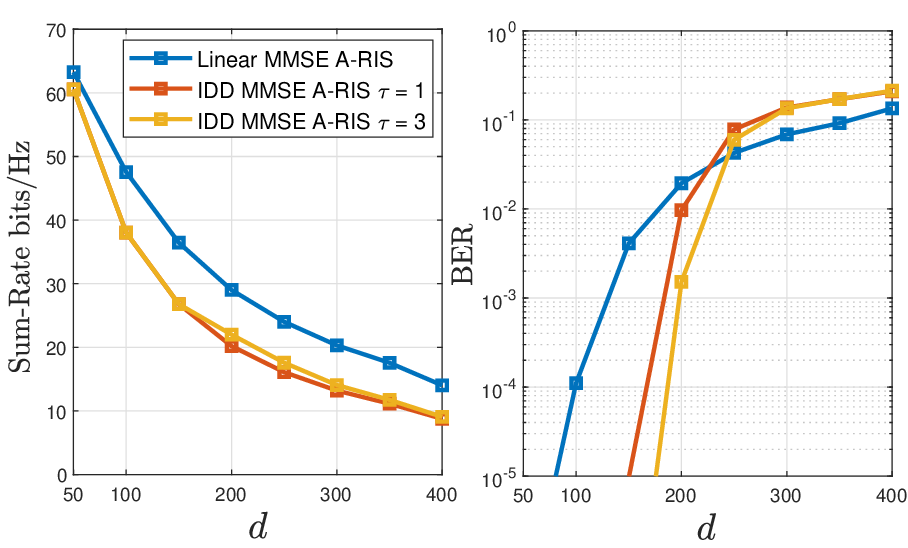}
    \caption{Uplink performance for $K = 12$, $M = 32$, $N = 64$, $\sigma_s^2 = -95$ dBm, $\sigma_v^2 = -95$ dBm and $P_T/K = 0$ dBm.}
    \label{fig:deploy_active}
    \vspace{-0.5cm}
\end{figure}

For the passive RIS scenario, $P_T/K = 6$ dBm, and the noise power is set to -100 dBm for $\sigma_s^2$ and 0 for $\sigma_v^2$. The corresponding results are shown in Fig. \ref{fig:deploy_passive}. In the active RIS scenario, $P_T/K = 0$ dBm, and the noise power is set to -95 dBm for both $\sigma_s^2$ and $\sigma_v^2$. The results for the active RIS are presented in Fig. \ref{fig:deploy_active}, where it is demonstrated that the sum-rate increases as the RIS moves closer to the AP and farther from the users. These findings corroborate the results discussed in Section III and emphasize the performance improvements achieved by the IDD in both scenarios.

\subsection{Number of users, AP antennas, and RIS elements}

\begin{figure}
    \centering
    \includegraphics[width=0.95\linewidth]{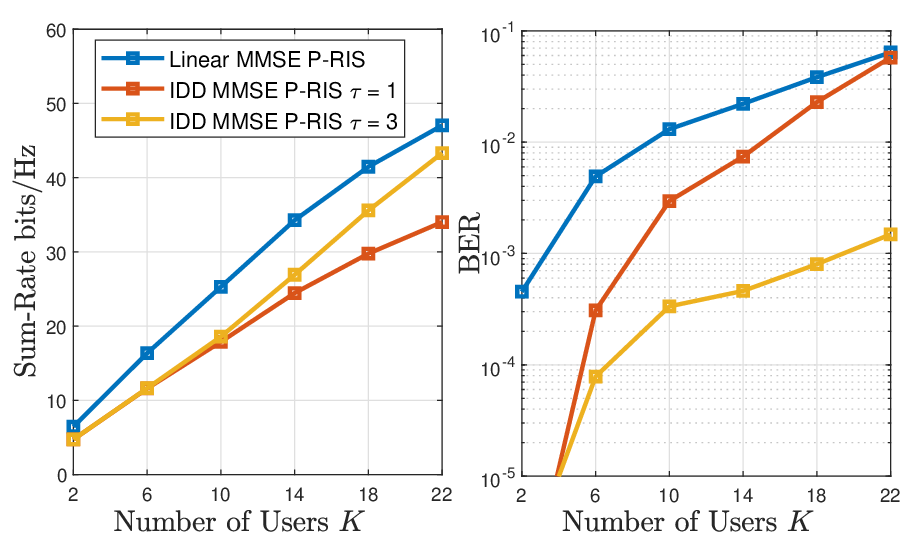}
    \caption{Performance for $M = 32$, $N = 64$, $\sigma_s^2 = -100$ dBm, $\sigma_v^2 = 0$ dBm and $P_T/K = 6$ dBm.}
    \label{fig:users}
    \vspace{-0.5cm}
\end{figure}

\begin{figure}
    \centering
    \vspace{-0.5cm}
    \includegraphics[width=0.95\linewidth]{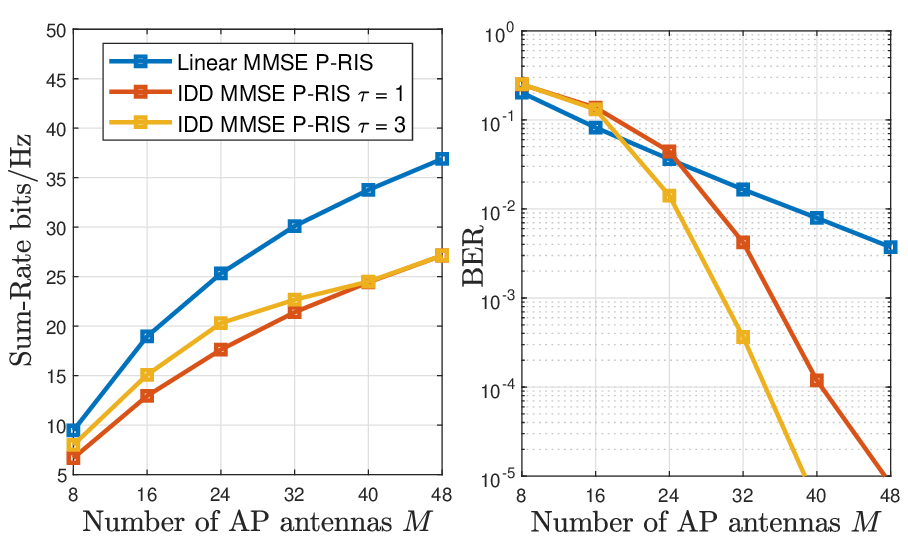}
    \caption{Performance for $K = 12$, $N = 64$, $\sigma_s^2 = -100$ dBm, $\sigma_v^2 = 0$ dBm and $P_T/K = 6$ dBm.}
    \label{fig:antennas}
\end{figure}

\begin{figure}
    \centering
    \includegraphics[width=0.95\linewidth]{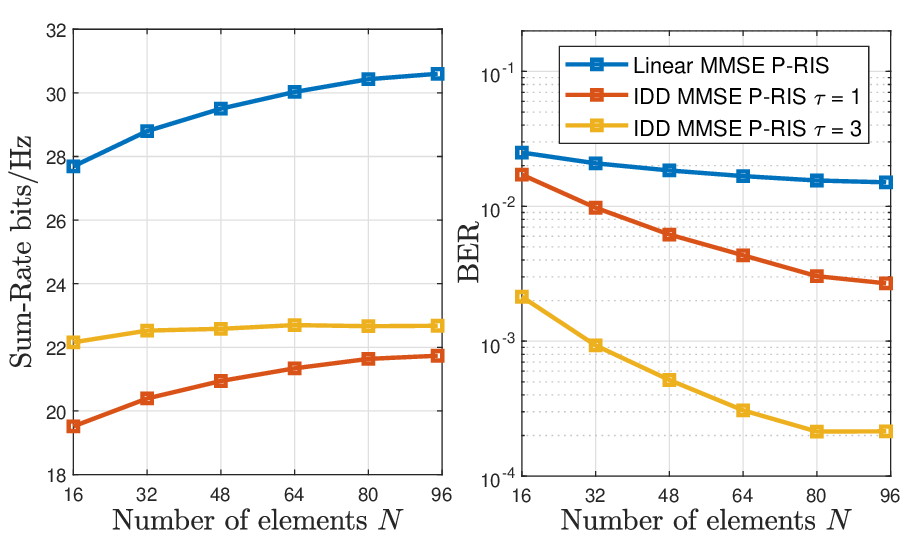}
    \caption{Performance for $K = 12$, $M = 32$, $\sigma_s^2 = -100$ dBm, $\sigma_v^2 = 0$ dBm and $P_T/K = 6$ dBm.   }
    \label{fig:RISelements}
    \vspace{-0.5cm}
\end{figure}

Instead of selecting random configurations, simulation results are presented with fixed element positions and user power. Specifically, $P_T/K = 6$ dBm is used, and the noise power is set to -100 dBm for $\sigma_s^2$ and 0 for $\sigma_v^2$. The system is configured with $K=12$ users, $M=32$ AP antennas, and $N=64$ RIS elements, which will remain constant unless explicitly stated otherwise. Additionally, the passive RIS is fixed at the position (0 m, 10 m). Performance variation is demonstrated by altering one parameter at a time (e.g., the number of users, antennas, or RIS elements), as shown in Fig. \ref{fig:users}, Fig. \ref{fig:antennas}, and Fig. \ref{fig:RISelements}.

\subsection{Fixed Position with Different Power Levels}
{
The system is configured with $K=12$ users, $M=32$ AP antennas, and $N=64$ RIS elements. The RIS is  positioned fixed at coordinates (0 m, 10 m). The results, presented in Fig. \ref{fig:fixed}, demonstrates significant improvements for the IDD MMSE P-RIS schemes compared to the Linear MMSE P-RIS, particularly with a higher number of IDD iterations. 

The overlap observed in the Sum-Rate occurs because the improvement brought by the extrinsic information, denoted as $\mathbf{L_C}$, in the IDD scheme only enhances the detector’s output under low SNR conditions. However, it is essential to note that, although $\mathbf{L_C}$ does not directly improve the sum-rate at the detector’s output, it still plays a significant role in updating $\mathbf{L_D}$, which is crucial for the decoder. This impact is reflected in the BER performance, which continues to benefit in higher SNR.

\begin{figure}
    \centering
    \includegraphics[width=0.95\linewidth]{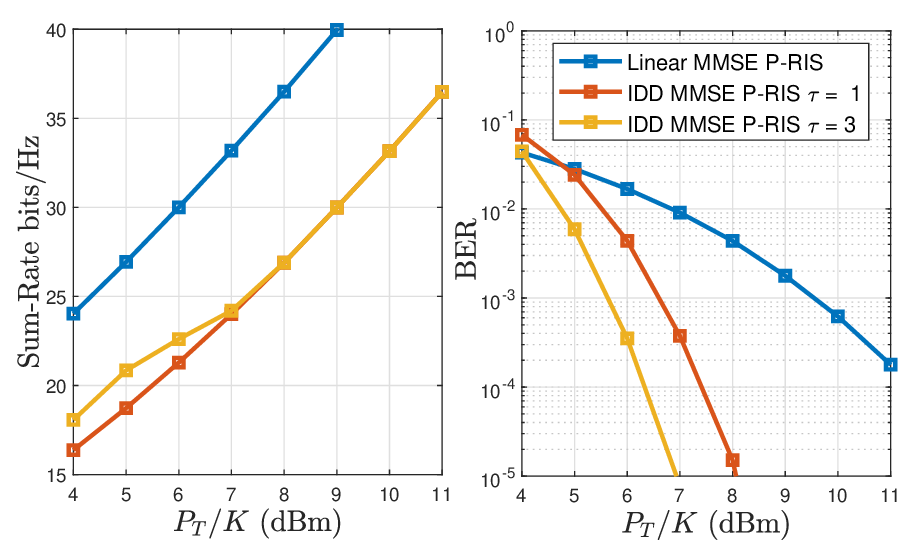}
    \caption{Performance for $K = 12$, $M = 32$, $N = 64$, $\sigma_s^2 = -100$ dBm and $\sigma_v^2 = 0$ dBm.  }
    \label{fig:fixed}
\end{figure}

\section{Conclusion}

This work presented a comprehensive study of how various parameters of an IDD RIS-assisted system impact its performance in terms of BER and sum-rate. Specifically, an analytical study is provided on the effects of deploying both active and passive RIS configurations in a SISO system. The results demonstrate that the analytical findings for the SISO system can be extended to a MIMO system with up to $K = 12$ users. Furthermore, it is shown that the IDD RIS system remains robust across a range of parameter values. In addition, it has been shown that the IDD RIS system remains robust across different parameter values.

\end{document}